\documentclass{moriond}

\def\Journal#1#2#3#4{{#1} {\bf #2}, #3 (#4)}

\def\PLB{{\em Phys. Lett.}  B}
\def\PRL{\em Phys. Rev. Lett.}
\def\PRD{{\em Phys. Rev.} D}

\def\JHEP{{\em J. High Energy Physics}}

\def\be{\begin{equation}}
\def\ee{\end{equation}}
\def\bea{\begin{eqnarray}}
\def\eea{\end{eqnarray}}

\def\ETm{$E_{\rm{T}}^{\rm{miss}}$}

\newcommand{\Photo}{}

\begin{document}
\vspace*{4cm}
\title{Thinking outside the beamspot: Other SUSY searches at the LHC (long-lived particles and
R-parity violation)}

\author{ H.W. Wulsin \\ 
for the ATLAS and CMS collaborations}

\address{Department of Physics, The Ohio State University, \\ 
191 West Woodruff Ave., Columbus OH, 43210, USA}

\maketitle\abstracts{
Supersymmetric particles that are long-lived or violate R-parity could evade many conventional searches for supersymmetry.  
This talk presents the latest results of searches for supersymmetry with long-lived
particles or R-parity violation performed by the ATLAS and CMS collaborations using $\sim$20
fb$^{-1}$ of proton-proton collisions at center-of-mass energy of 8 TeV delivered by the LHC.
}

\section{Overview}
Supersymmetry (SUSY) is an attractive theory because it could provide solutions to two big unsolved mysteries in particle physics:  the identity of a dark matter particle and the stability of the Higgs mass against quantum corrections at very high energy scales.  
Many searches for SUSY have been performed, but no evidence of its existence has yet been found.  
Supersymmetric particles with long lifetimes could evade many conventional SUSY searches, which typically require the physics objects of interest to have a transverse impact parameter less than about a millimeter.  R-parity violating SUSY particles could also evade many SUSY searches, which commonly require a large amount of missing transverse energy \ETm.  
To cover these gaps, dedicated searches that target long-lived particles and R-parity violation have been performed with data collected by the ATLAS~\cite{ATLASDet} and CMS~\cite{CMSDet} experiments, in most cases using $\sim$20 fb$^{-1}$ of $\sqrt{s}$ = 8 TeV proton-proton collisions delivered by the LHC in 2012.  No significant excesses above the background expectation are observed, and limits are placed on a variety of models.

The searches presented here cover a wide variety of signatures, including jets, leptons, photons, and tracks.
These signatures depend on whether the supersymmetric particle is neutral or charged, and
in what region of the detector it decays: the tracker, calorimeters, muon system, or outside the
detector entirely.  In general there are few standard model backgrounds that can mimic these signatures, but these
searches have to confront unusual backgrounds from sources such as detector noise, cosmic rays,
and reconstruction failure modes.  The simulation is typically unable to describe these processes
accurately, so the backgrounds are usually estimated with data-driven techniques.  

\section{Neutral long-lived particles decaying to jets, leptons, or photons}
\begin{figure}[h!]
\begin{minipage}{0.5\linewidth}
\centerline{\includegraphics[width=1.0\linewidth]{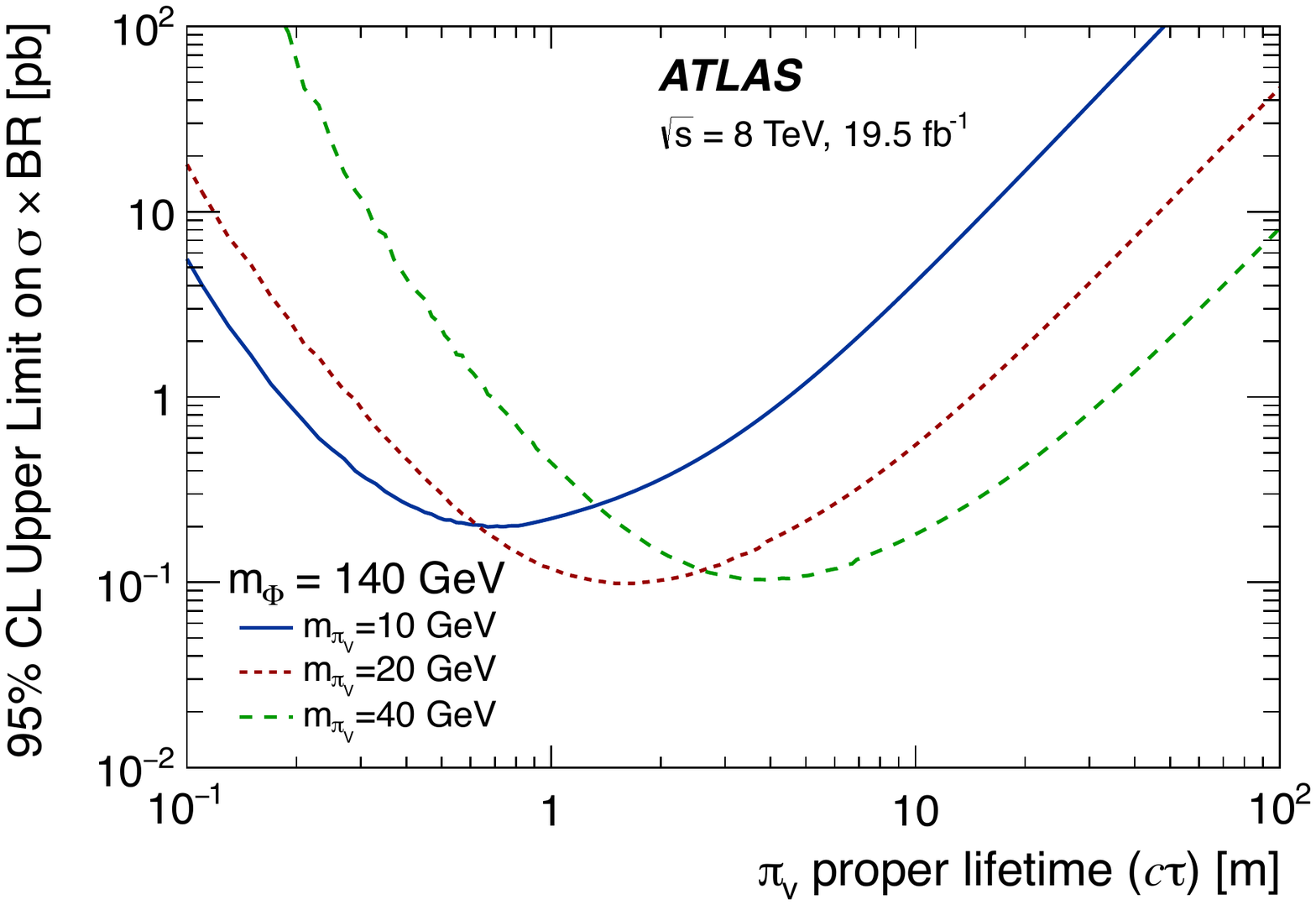}}
\centerline{\includegraphics[width=0.95\linewidth]{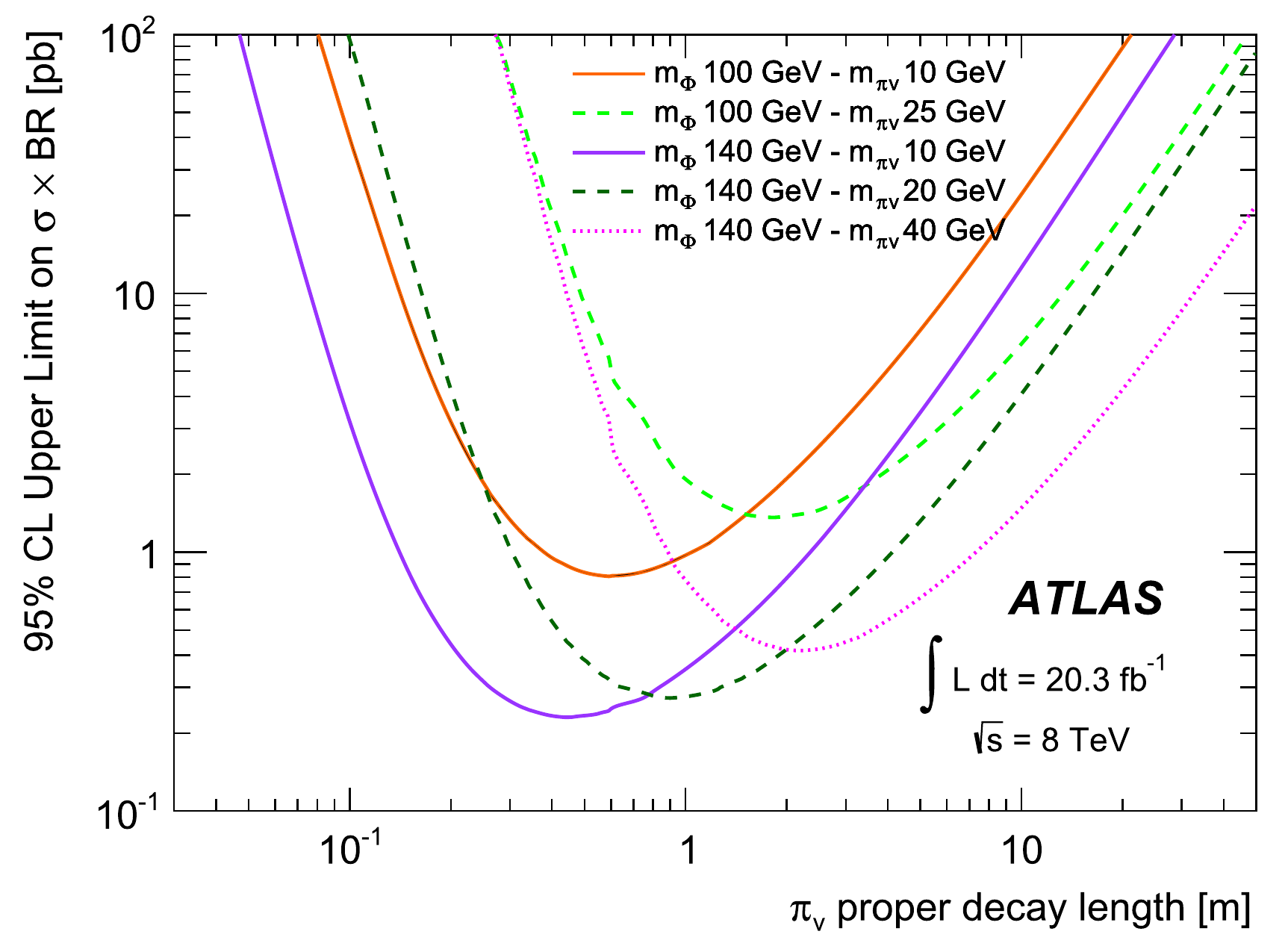}}
\end{minipage}
\hfill
\begin{minipage}{0.5\linewidth}
\centerline{\includegraphics[width=0.8\linewidth]{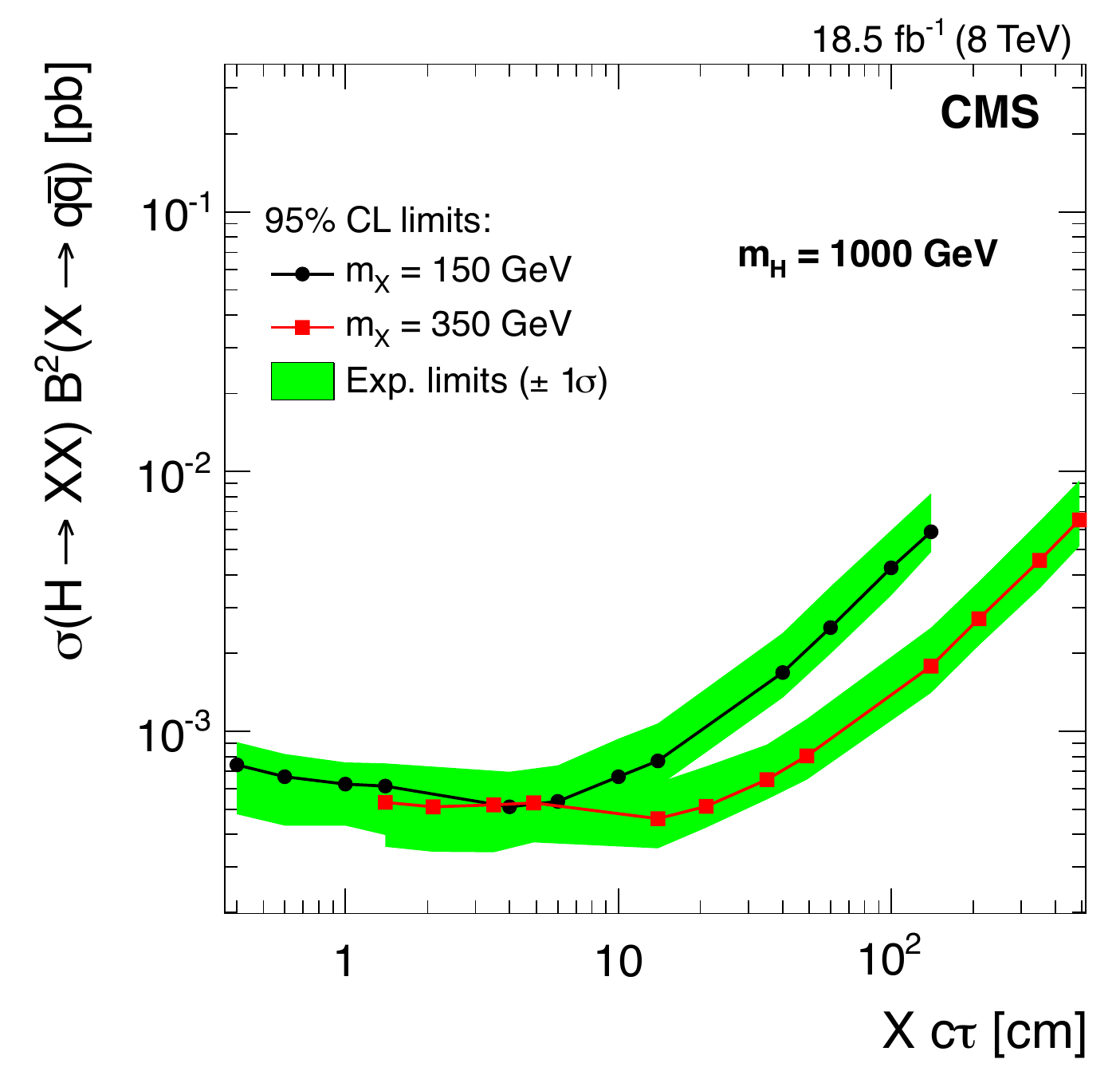}}
\centerline{\includegraphics[width=0.8\linewidth]{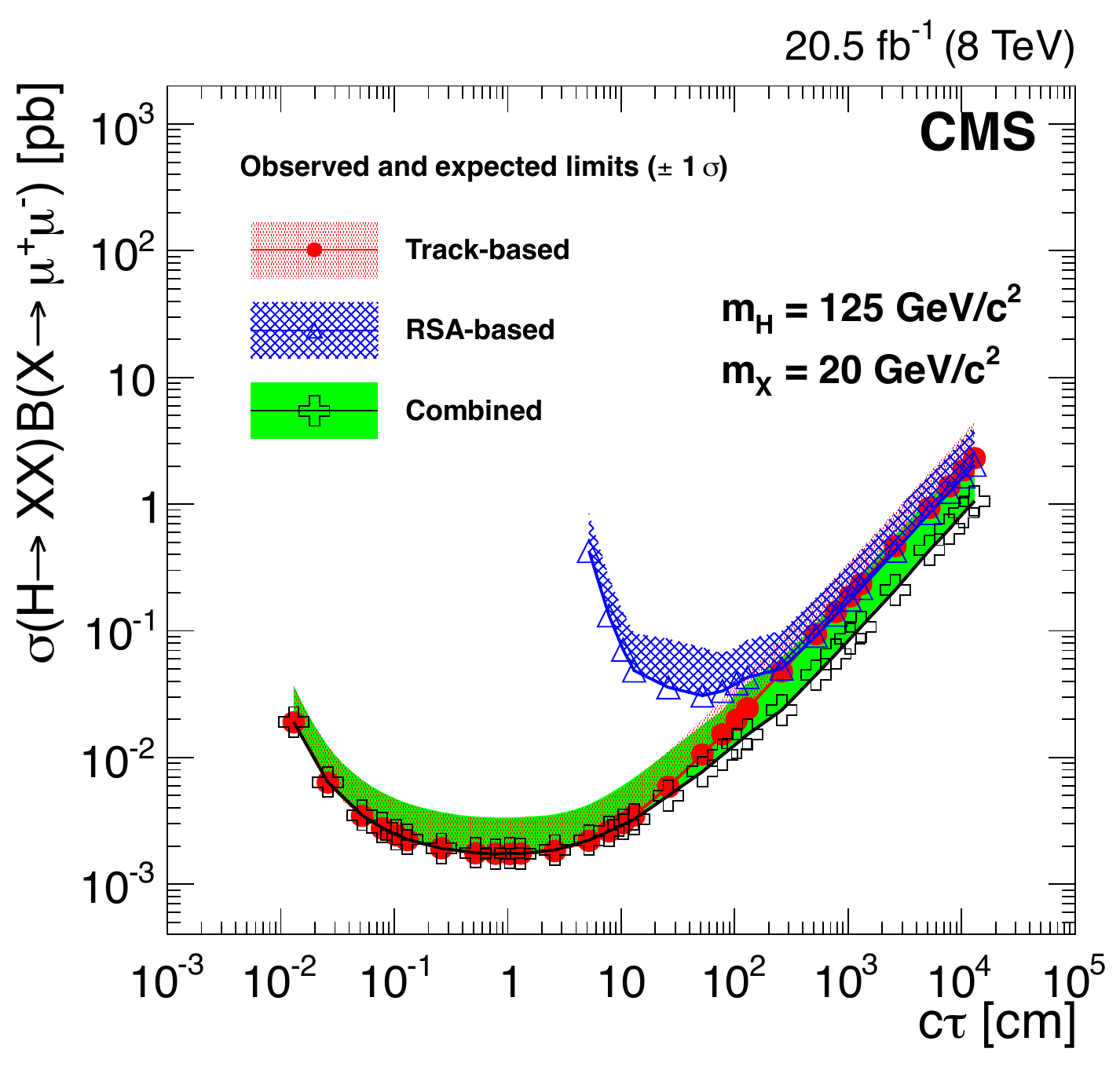}}
\end{minipage}
\caption[]{Limits are placed on hidden valley models with a long-lived neutral particle, based on the signature of two jets from a common vertex (upper left~\cite{displJetsATLAS} and upper right~\cite{dispJetsCMS}), two trackless jets (lower left~\cite{tracklessJetsATLAS}), and two opposite-sign leptons from a common vertex (lower right~\cite{displSTAMuonsCMS}).}  
\label{fig:hiddenValley}
\end{figure}

A search for displaced dijets~\cite{dispJetsCMS} requires two jets, each with at least
one track, originating from a common displaced vertex. There are many ways to produce such a signature, including 
a single displaced jet with final state radiation, a decay to 3 or more jets, or a decay
to leptons, which are included as jets.  
The dominant background is from QCD multijets, which is reduced by placing requirements on the
number of prompt tracks and the fraction of energy they carry, as well as on a likelihood
discriminant that combines four variables associated with the vertex and cluster information.
The remaining background is estimated by using uncorrelated data sideband control regions.
Two events are observed in a loose search region and one event is observed in a
tight search region, both consistent with expectations.
Limits are set on hidden valley models and on RPV SUSY for decay lengths of millimeters to
several meters; an example is given in Fig.~\ref{fig:hiddenValley}.  

If the long lived particle decays in the hadronic calorimeter, it would
produce the signature of a trackless jet, a narrow jet with few associated tracker tracks
and very little energy deposited in the electromagnetic calorimeter. A search for such particles~\cite{tracklessJetsATLAS} estimates the dominant QCD multijets background from the data. In the search region, 24 events are observed, consistent with the
expectation, and limits are set on a hidden valley model in terms of the decay length of the valley pion
ranging from tens of centimeters to tens of meters; an example is given in Fig.~\ref{fig:hiddenValley}.  

Another search~\cite{displJetsATLAS} selects events with two displaced vertices close to a jet,
each with at least five tracks, with the vertices either in the tracker or in the muon system.  
The backgrounds from fake vertices are estimated from data control regions, and are predicted to be two events or less for five different event topologies.  
Limits are set on a hidden valley scalar boson $Z'$ and on stealth SUSY, with sensitivity to decay lengths between tens of centimeters and tens of meters; an example is given in Fig.~\ref{fig:hiddenValley}.  

There are also models that predict a long-lived particle that decays predominantly into light leptons,
producing collimated jets of electrons, muons, or both. A 
search~\cite{leptonJets} requires two such lepton jets in the event, each of which is
isolated from other tracks, and also separated from each other in the azimuthal angle. The
observation is consistent with the expectation, and limits are placed on a dark photon model with a
maximum sensitivity to decay lengths of several centimeters.

A search for a long-lived particle that decays to two opposite-sign, same-flavor leptons~\cite{displLeptonsCMS} suppresses the large contribution from Drell-Yan processes by
placing cuts on the transverse impact parameter of the two leptons. 
The additional background is estimated
with a data control region in which dilepton momentum is opposite the flight direction from the
primary to the secondary vertex. 
Zero events are
observed in the control region as well as in the signal region, and limits are set on a hidden valley
model and on RPV SUSY. This search using inner-tracker tracks has recently been combined with a
search that uses only muon tracks~\cite{displSTAMuonsCMS}; an example is given in Fig.~\ref{fig:hiddenValley}.  

ATLAS has performed searches for displaced vertices~\cite{displVtxATLAS} within 30 cm of the beamspot in three dilepton channels and four multitrack channels. 
Vertices close to any detector material are vetoed in order to suppress the background from nuclear
interactions.  
In all seven search channels, the main backgrounds originate from unrelated tracks or leptons
that cross at a large angle to give a displaced vertex. In the multitrack channels, the vertex is required to have an invariant mass of at least 10 GeV and at least 5 tracks. 
The backgrounds are predicted to be less than one event, and no
events are observed in any of the search channels.  Limits are placed on split SUSY, RPV, and general GMSB.  

A CMS search~\cite{displSUSY} motivated by a Displaced SUSY model selects events with an electron and a muon of opposite sign, each displaced, but not necessarily originating from a common vertex.
This search
defines three exclusive signal regions based on the lepton transverse impact parameter, in the range of 0.2 mm to 2 cm.  In the first region there are 19 observed events, and in the other two regions there are no observed events. These observations are 
consistent with the background predictions, and limits are placed on the stop lifetime and mass with
a maximum sensitivity for a stop lifetime of about 2 cm.

If the long-lived particle decays to a photon it can be identified by the fact that the photon does
not point back to the primary vertex and that it arrives late at the calorimeter. A search for displaced/delayed photons~\cite{displPhotonsATLAS} exploits the pointing and timing resolution of the ATLAS detector's liquid argon calorimeter. 
The signal region is defined as two high-energy photons with large \ETm, and the search is
performed in the two-dimensional plane of the longitudinal impact parameter and calorimeter arrival time.  A low-\ETm control
region is used to estimate the background. Limits are set on the GMSB model with a maximum sensitivity to lifetimes of
about 2 ns.

\section{Charged long-lived particles}
If the long-lived particle is charged and has sufficiently low kinetic energy, it may come to rest
in the detector. After some time, it may then decay. A CMS search for stopped particles~\cite{stoppedPartCMS} selects events with a
calorimeter cluster that is asynchronous with the proton-proton collisions. 
The search sample corresponds to 281
hours of trigger livetime. The backgrounds in this search arise from processes that are uncorrelated with the
proton-proton collisions, such as beam halo muons, cosmic rays, and HCAL noise. Ten events are
observed, consistent with the expectation, and limits are placed on the gluino and stop masses for
over 13 orders of magnitude.  A similar search has been performed by ATLAS~\cite{stoppedPartATLAS}.  

\begin{figure}
\begin{minipage}{0.5\linewidth}
\centerline{\includegraphics[width=0.8\linewidth]{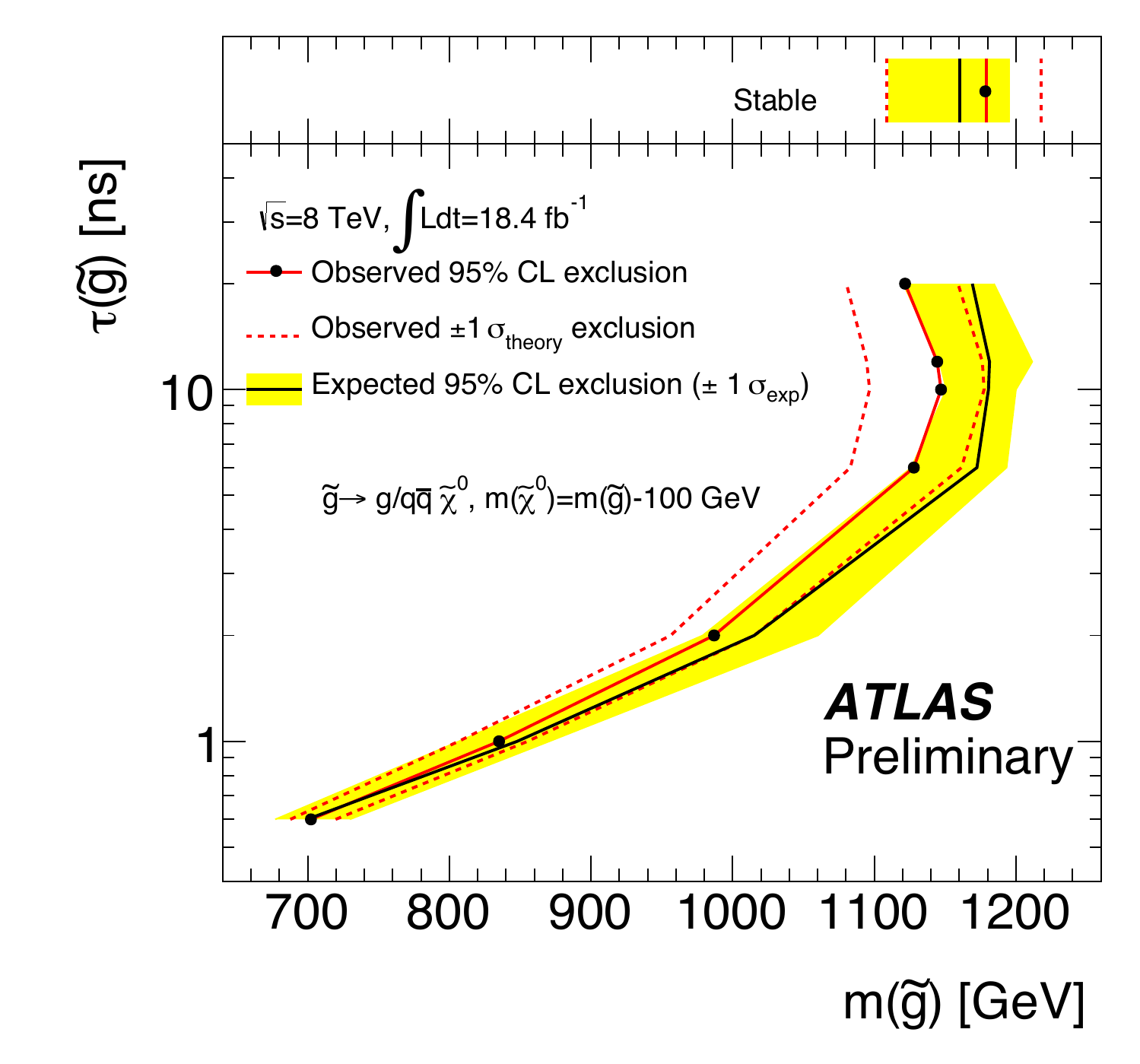}}
\end{minipage}
\hfill
\begin{minipage}{0.5\linewidth}
\centerline{\includegraphics[width=0.8\linewidth]{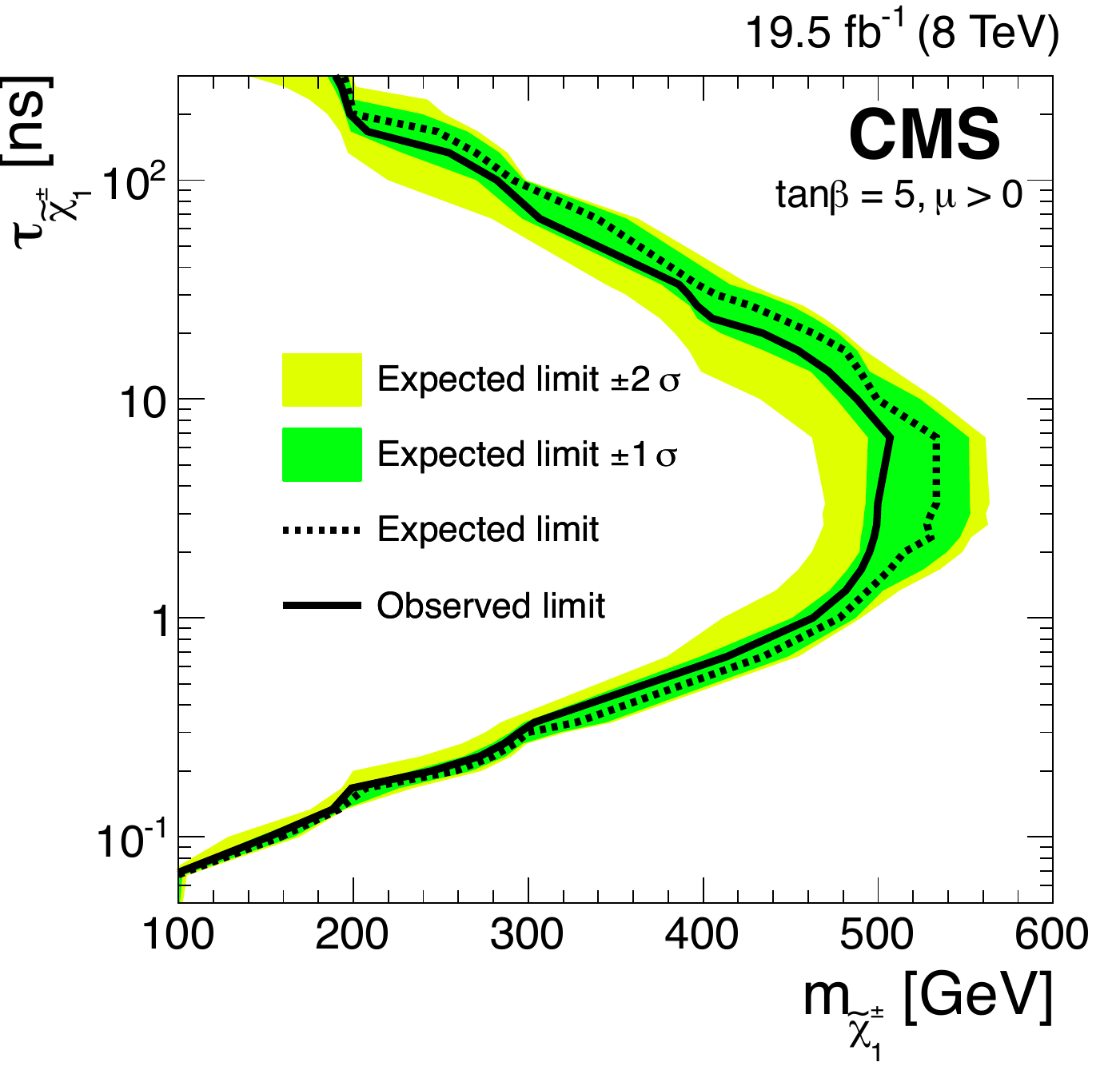}}
\end{minipage}
\caption[]{Limits on charged metastable particles are placed using dE/dx information~\cite{HSCPShortTrkATLAS} (left) and by identifying disappearing tracks~\cite{disappTrksCMS} (right).}
\label{fig:shortTrks}
\end{figure}

A long-lived charged particle can also be identified if it does not come to rest in the detector and
escapes the detector, based on the fact that it is moving slowly. A search for such particles~\cite{HSCPATLAS} identifies tracks with large
ionization energy loss dE/dx in the pixel tracker or late timing measurements in the calorimeter
or the muon system. Based on these measurements, the mass of the particle is reconstructed, which for signal is much larger than for Standard Model particles. The main background is
from muons that have mismeasured dE/dx or timing information, and it is estimated by
sampling the relevant distributions from data control regions. No excess is observed over the
expectations, and limits are set on GMSB sleptons, LeptoSUSY squarks and gluinos, charginos, and R-hadrons.  A related search~\cite{HSCPMultiATLAS} measures dE/dx from multiple
subsystems and places limits on particles with a charge between 2 and 6 times the charge of the electron.  
A search for metastable charged particles~\cite{HSCPShortTrkATLAS} uses dE/dx measurements from the ATLAS pixel tracker and 
places limits on R-hadrons with lifetimes between 1 and 10
ns, stable gluinos (see Fig.~\ref{fig:shortTrks}), and stable and metastable charginos.  
Searches for particles with anomalous dE/dx measurements have also been performed by CMS~\cite{HSCPCMS}.  

A search for charged long-lived particles is also performed using a disappearing track signature, that of
a high-$p_T$ isolated track with several missing hits in the outer layers of the tracker and little
energy deposited in the calorimeters~\cite{disappTrksCMS}. The backgrounds in this search are estimated with tag and probe methods, and arise from reconstruction
failure modes: unidentified electrons or muons, tracks with mismeasured momenta, or fake tracks. 
Two events are observed, consistent with
the expectations, and limits are placed on the chargino mass and lifetime, with maximum sensitivity
to lifetimes of several nanoseconds, as shown in Fig.~\ref{fig:shortTrks}.  A similar search has been performed by ATLAS~\cite{disappTrksATLAS}.  

\section{R-parity violating prompt decays}
A search for RPV SUSY in multijet final states~\cite{RPVMultijetATLAS} uses two strategies, a jet-counting analysis that looks for an excess of $\geq$6 or $\geq$7 jet events, with 0-2 b-jets, and an analysis based on the total jet mass, which is defined as the scalar sum of masses of the four leading large-radius jets, which would be formed from accidental substructure.  The background is estimated from signal-depleted data control regions, in the jet-counting analysis by extrapolating with a scale factor from simulation, and in the total jet mass analysis by constructing the total jet mass from single jet mass templates.  No excess above the prediction is observed, and limits are placed in the plane of gluino-neutralino masses.  

Another RPV search~\cite{RPVLFVATLAS} looks for a lepton flavor violating resonance of $e^\pm\mu^\mp$, $e^\pm\tau^\mp$, or $\mu^\pm\tau^\mp$.  The search places lower limits on the $\tau$-sneutrino mass between 1.7 - 2.0 TeV and on the Z' vector boson between 2.2 - 2.5 TeV, depending on the channel.  

Finally, a search~\cite{RPVBLStopATLAS} is performed for pair-produced scalar tops that decay via an R-parity-violating coupling to a final state with two opposite-sign leptons of any flavor and 2 b-jets.  The signal signature is two lepton-$b$-quark resonances with a similar mass.  Limits are placed on the scalar top mass between 0.5 and 1 TeV in the framework of a $B-L$ extension of the Standard Model.  

\section{Reinterpretations}  
There are two new reinterpretations of searches that place constraints on long-lived particles. ATLAS reinterprets a prompt SUSY search for jets and \ETm and sets mass limits on gluinos with lifetimes of
up to over a microsecond~\cite{reinterpretGluinoATLAS}. A CMS analysis~\cite{pMSSMCMS} parameterizes the signal efficiencies of its searches for
heavy stable charged particles, allowing a recasting in terms of the pMSSM and other models.

\section*{Acknowledgments}
I thank the members of the ATLAS and CMS collaborations for producing the results presented in this
talk,
the CERN accelerator division for the excellent operation of the LHC, and the many funding agencies
that support these experiments.

\end{document}